\documentclass[mathleft
]{an}
\usepackage{graphicx}
\usepackage{times}
\def\dss {{$\delta$~Scuti~stars}}
\def\gds {{$\gamma$~Doradus~stars}}
\def\bceph {{$\beta$~Cephei}}

\def\msun {{\mathrm{M}_\odot}}

\def\Os {{\Omega_{\rm s}}}
\def\Sm {{S_m}}

\def\dr {{\rm d}r}
\def\dth {{\rm d}\theta}
\def\nur {{\nu_{\rm ref}}}

\newcommand{\eqn} [1] {
\begin{equation}
#1
\end{equation}}

\begin{document}

\Pagespan{789}{}
\Yearpublication{2006}%
\Yearsubmission{2005}%
\Month{11}%
\Volume{999}%
\Issue{88}%

\title{On the use of rotational splitting asymmetries to probe the internal rotation
profile of stars. Application to \bceph\ stars}

\author{J.C.~Su\'arez\inst{1}\fnmsep\thanks{Corresponding author:
  \email{jcsuarez@iaa.es}\newline}
\and L.~Andrade\inst{2}
\and M.J.~Goupil\inst{3}
\and E.~Janot-Pacheco\inst{2}
}
\titlerunning{Analysis of internal rotation profile of stars}
\authorrunning{J.C. Su\'arez et al.}
\institute{
Instituto de Astrof\'{\i}sica de Andaluc\'{\i}a (CSIC), Rotonda de la Astronom\'{\i}a
S/N, Granada, 3004, Granada, Spain.
\and 
Instituo de Astronomia, Geof\'{\i}sica e Ci\^encias Atmosf\'ericas da Universidade de
S\~ao Paulo (IAG-USP), Rua do Mat\~ao, 1226, S\~ao Paulo, Brazil
\and 
LESIA, Observatoire de Paris-Meudon, UMR8109, Meudon,France.}

\received{30 May 2005}
\accepted{11 Nov 2005}
\publonline{later}


\abstract{
  Rotationally-split modes can provide valuable information about the internal
rotation profile of stars. This has been used for years to infer the internal
rotation behavior of the Sun. The present work discusses the potential additional
information that rotationally splitting asymmetries may provide when studying
the internal rotation profile of stars. 
 We present here some preliminary results of a method, currently under
 development, which intends: 1) to understand the variation of the rotational
 splitting asymmetries in terms of physical processes acting on the angular
 momentum distribution in the stellar interior, and 2) how this information can
 be used to better constrain the internal rotation profile of the stars.
 The accomplishment of these two objectives should allow us to better use asteroseismology
 as a test-bench of the different theories describing the angular momentum 
 distribution and evolution in the stellar interiors.}

\maketitle

\section{Introduction}

The study of internal rotation of stars is one of the main issues in stellar 
physics. Rotation is present in almost all the stars, and it interacts with other
physical processes acting in the stellar interior. In particular, understand the 
transport of angular momentum in the interior of stars is crucial to correctly and 
precisely describe the evolution. Turbulence, meridional circulation, mixing of 
elements responsible of different $\mu$ gradients during evolution, dynamo effects
due to the presence of magnetic fields, etc., are some of the physical phenomena and processes
affected by rotation (see e.g. Goupil et al. 2005, Goupil 2009, and Goupil \& Talon 2009, for a
review on this topic).

Nowadays, it is possible to probe the internal structure of stars thanks to 
asteroseismology. In what regards rotation, progress has been made, during
the last decades, on the knowledge about the rotation-pulsation interaction.
Up to now, this problem has been tackled using the perturbation techniques 
to compute the stellar oscillations. These methods (see e.g. 
Dziembowski \& Goode 1992; Soufi et al. 1998; Su\'arez, Goupil \& Morel 2006) 
are only valid for slow-to-moderate rotators (see a review on the effects of 
rotation on stellar p modes by Goupil 2009). For faster rotators, the distortion 
of the stellar structure due to the centrifugal is too large and invalids the 
perturbation techniques. Non-perturbative approaches must thus be considered 
(Ligni\`eres et al. 2006; Reese et al. 2006).

The present work consider slow-to-moderate rotators, i.e. those for which the
parameters  $\epsilon=\Omega/(G\,M/R^3)^{1/2}$ and $\mu=\Omega/\nu_{n,\ell}$ are small, 
i.e. the stellar structure is not significantly deformed by the centrifugal force, and oscillation
frequencies are much larger than the angular rotation rate, respectively. The failure of the
perturbative approach comes first for high radial-order frequency modes. We restrict thus this
study to rotating stars showing oscillations in a relatively low-order frequency domain (low g-
and p modes, or mixed modes), like some \dss, \gds, solar-like stars and some massive stars like
\bceph.

The heterogeneity of internal structures (chan\-ging with evolution and different from
 one star to another) and processes therein, are strong
arguments to assume non-uniform rotation. It is thus necessary to take into account possible
variations (in the radial or angular directions) of the angular momentum transport, and thus in the
shape of the internal rotation profile. To do so, we use the oscillation code {\sc filou}
(Su\'arez 2002; Su\'arez \& Goupil 2008) which corrects the oscillation frequencies for up to 
second-order effects of rotation (including near degeneracy effects) in presence of radial
differential rotation. The study of radial differential rotation has also been used for the
asteroseismic studies, e.g. Casas~(2006, 2009), Fox Machado ~(2006), Bruntt et al.~(2007a, 2007b);
Su\'arez et al.~(2005a), Su\'arez et al.~(2006b); Su\'arez et al.~(2007), for \dss, Rodr\'{\i}guez
et al.~(2006a, 2006b), Uytterhoeven et al.~(2008), Moya et al.~(2005), Su\'arez et al.~(2005b), for
\gds, or Su\'arez et al.~(2009) for \bceph\ stars.

This technique is applied in the present work to analyze the asymmetries of the mode splittings due
to rotation. In particular, we are interested in examining the behavior of the rotational splitting
and their asymmetries for different mode types (g and p), in presence of radial differential
rotation, i.e. to understand physically how variations of the internal rotation profile affect the
splitting asymmetries.

\section{Rotational splittings\label{sec:splittings}}

Rotational splittings are being used so far, using inversion techniques, to determine the internal
rotation of stars. The most succeeded example of this is the Sun, for which, the internal rotation
profile is known precisely thanks to helioseismic inversions (Thompson et al. 1996).

In the framework of the perturbation theory, rotational splittings correspond to the first-order in
$\Omega$ (rotational frequency) correction to the oscillation frequency $\omega_{n,\ell,m}$. In that
case, the Coriolis acceleration ($2\Omega\times{\bf v}$) dominates, and the frequency correction
term can be written as
\eqn{\omega_{1,m}= \int_0^R\int_0^\pi K_m(r,\theta)\,\Omega(r,\theta)\dth\dr\label{eq:defw1}}
where $K_m(r,\theta)$ is the rotational splitting kernel, and $R$ the stellar radius. We use this
expression to define the generalized splitting $\Sm$ as
\eqn{\Sm = \frac{\omega_{1,m}-\omega_{1,-m}}{2m}\label{eq:defSm}}
For the sake of simplicity, we assume here shellular rotation, i.e. $\Omega(r)=\Os[1+\eta_0(r)]$, so
that the splitting becomes independent of $m$, and its kernel can be written as
\eqn{K(r)= \frac{\xi_r^2-2\xi_r\xi_h+(\Lambda-1)\xi_h^2}{\xi_r^2+\Lambda\xi_h^2}
           r^2\rho_0\label{eq:defKr}}
where $\xi_r$ and $\xi_h$ correspond to the radial and horizontal components of the eigenfunction,
respectively. The density of the star (unperturbed) is represented by $\rho_0$, the radial
distance by $r$, and $\Lambda = \ell(\ell+1)$. The denominator of this equation is generally known
as the mode inertia.
When rotation is assumed uniform, the rotational splitting simplifies to 
\eqn{S = \Omega\int_0^R K(r)\dr\label{eq:defSunif}}
For a shellular rotation, the rotational splitting can be written in the form given by 
Su\'arez et al.~(2006a), as
\eqn{S = m\,\Os\, (C_{\rm L}-1-J_0)\label{eq:Srotdif}}
where $C_{\rm L}$ is the Ledoux constant, and $J_0$ is a $\eta$ dependent integral (which is null
for uniform rotation), defined by
\eqn{J_0 =\frac{1}{I_0}\int_0^R \eta_0(r)[y_{01}^2+\Lambda\,z_{01}^2-2y_{01}z_{01}-z_{01}^2]\,
           \rho_0\,r^4\,\dr\,.\label{eq:defJ0}}
with $I_0$ being the mode inertia previously defined in Eq.~\ref{eq:defKr}.
\section{Rotational splitting asymmetries\label{sec:asym}}

It often observed that rotational splitting are not perfectly symmetric (with respect to 
axisymmetric modes). Mathematically, asymmetries of splitting can be defined, in its generalized
form, as
\eqn{A_m = \omega_{-m}+\omega_{+m}-2\omega_{0}\,.\label{eq:defasym1}}
Doing some simple algebra, it can be shown that splitting asymmetries are dependent only
upon second-order (in $\Omega$) terms, particularly on
\eqn{A_m = m^2\,X_2\,\frac{\Os^2}{\omega_0}\label{eq:defasym2}}
where expressions for $X_2$ (the Saio form) can be found in e.g. Su\'arez et al. 2006a or
Goupil~(2009) and references therein. Then, it thus possible to construct the kernel for
the splitting asymmetries, $K_2$ such as
\eqn{A_m = m^2\int_0^R \Omega^2(r)K_2(r)r^2\rho_0\dr\label{eq:defasym3}}
which is implicitly dependent on the rotation profile $\eta_0$. In this work we are attempting to
examine $K_2$ in detail. To do so, it is necessary to investigate the behavior of $A_m$ for
different mode types and its sensitivity to variations of the rotation profile in different zones
of the stellar interior. 
 
\begin{figure}
   \includegraphics[width=8cm]{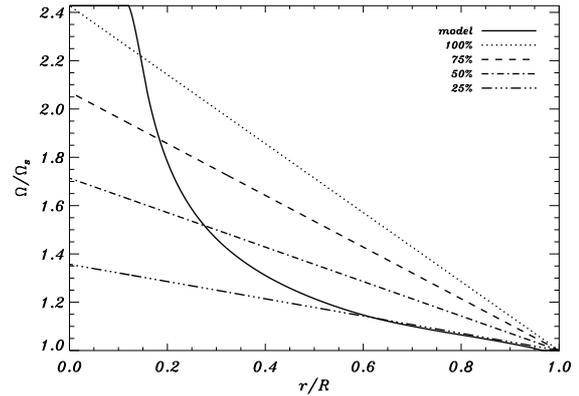}
    \caption{Two types of rotation profile: linear and shellular (local conservation of 
             angular momentum).The linear profiles were constructed keeping the same rotational
             frequency at the stellar surface, but varying it at the core, approximately at
             100\%, 75\%, 50\%, and 25\% of the original one $\nu_\Omega=2.42$. These are compared
             with a shellular type rotation profile with the same rotational frequency at the
             surface.}
\label{fig:rotprof1}
\end{figure}
\section{Probing rotation profiles using splitting asymmetries\label{sec:asym}}
\begin{figure*}
   \includegraphics[width=8cm]{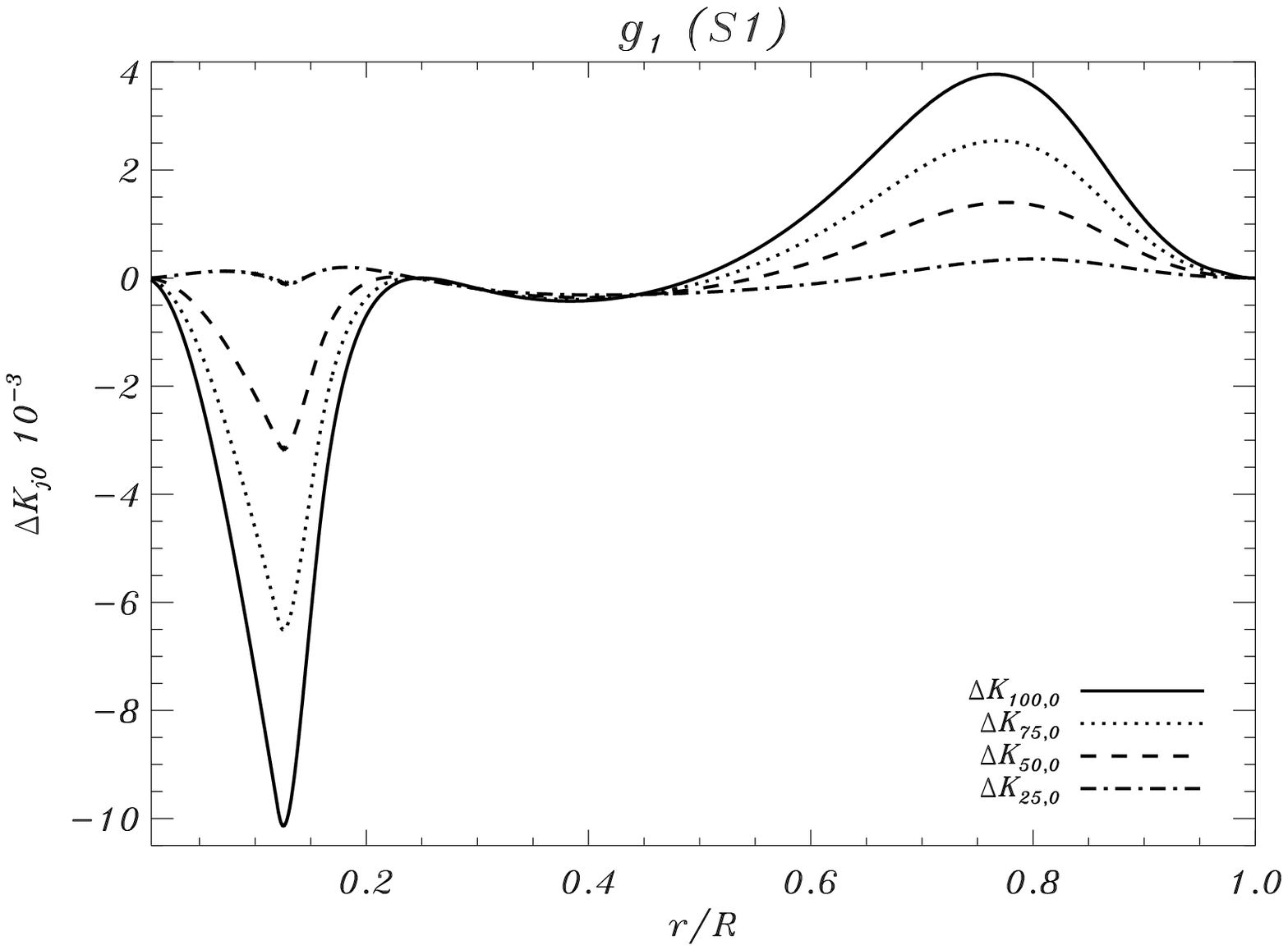}
   \includegraphics[width=8cm]{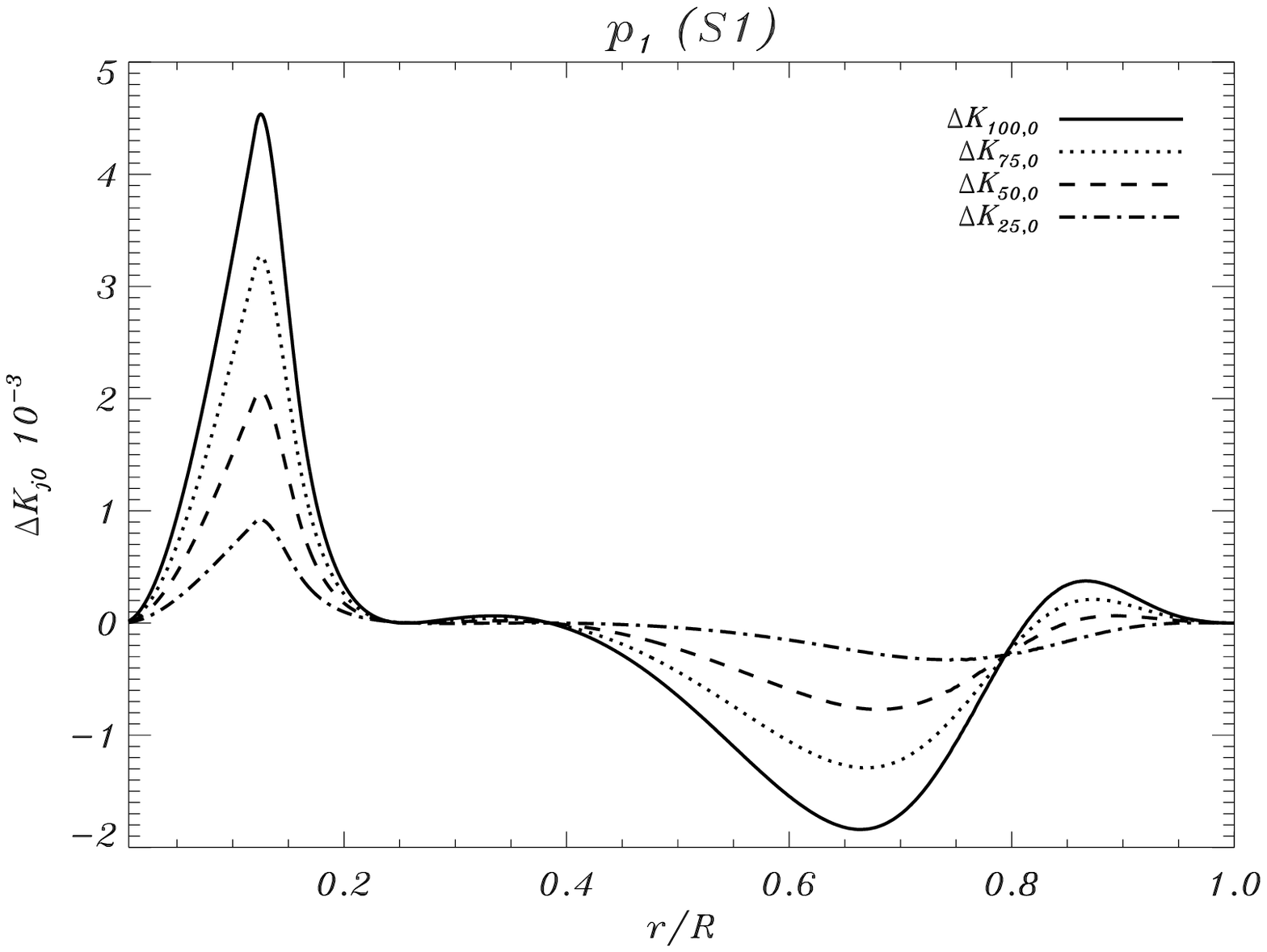}  
    \includegraphics[width=8cm]{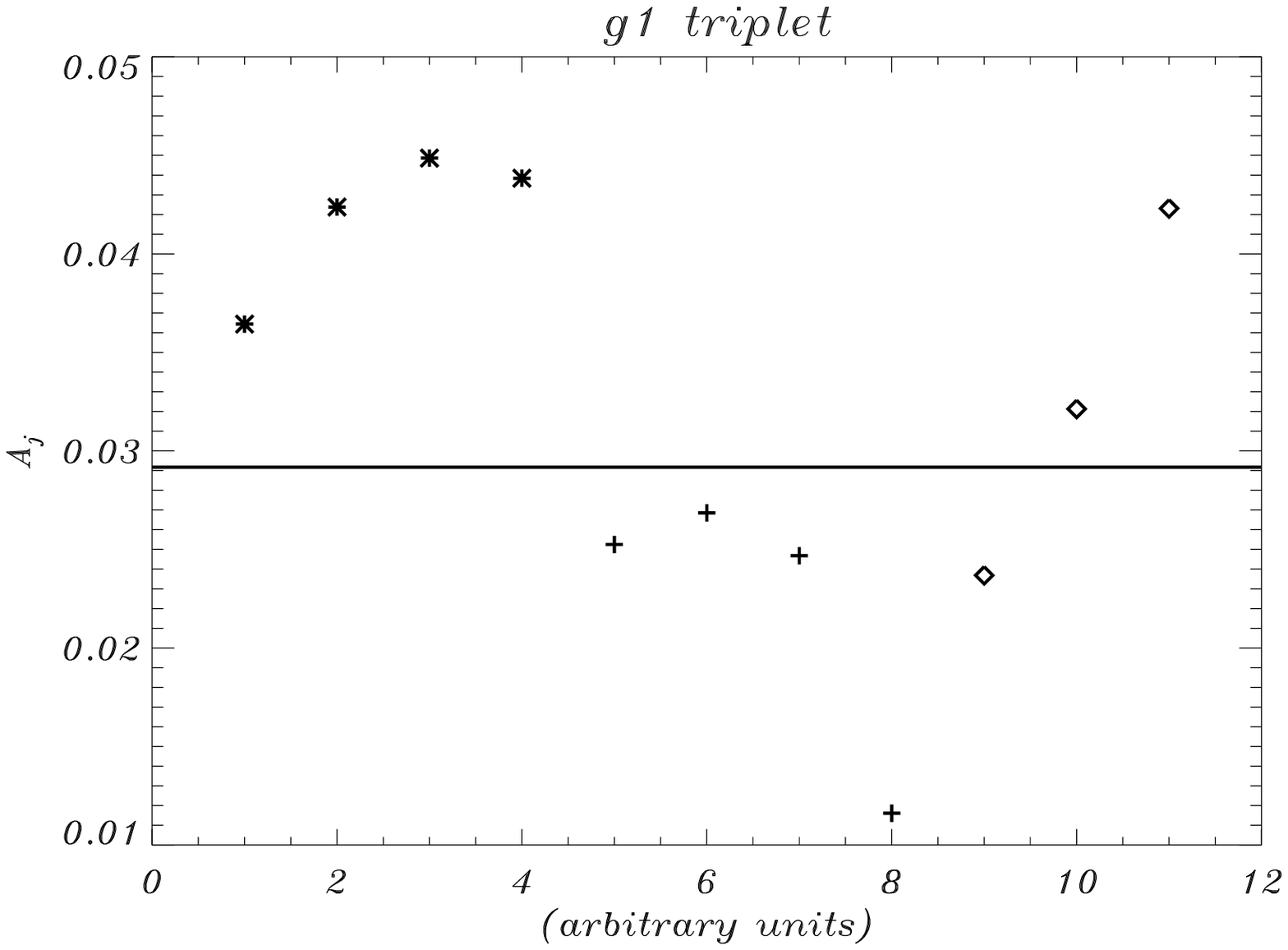}\hspace{1cm}   
    \includegraphics[width=8cm]{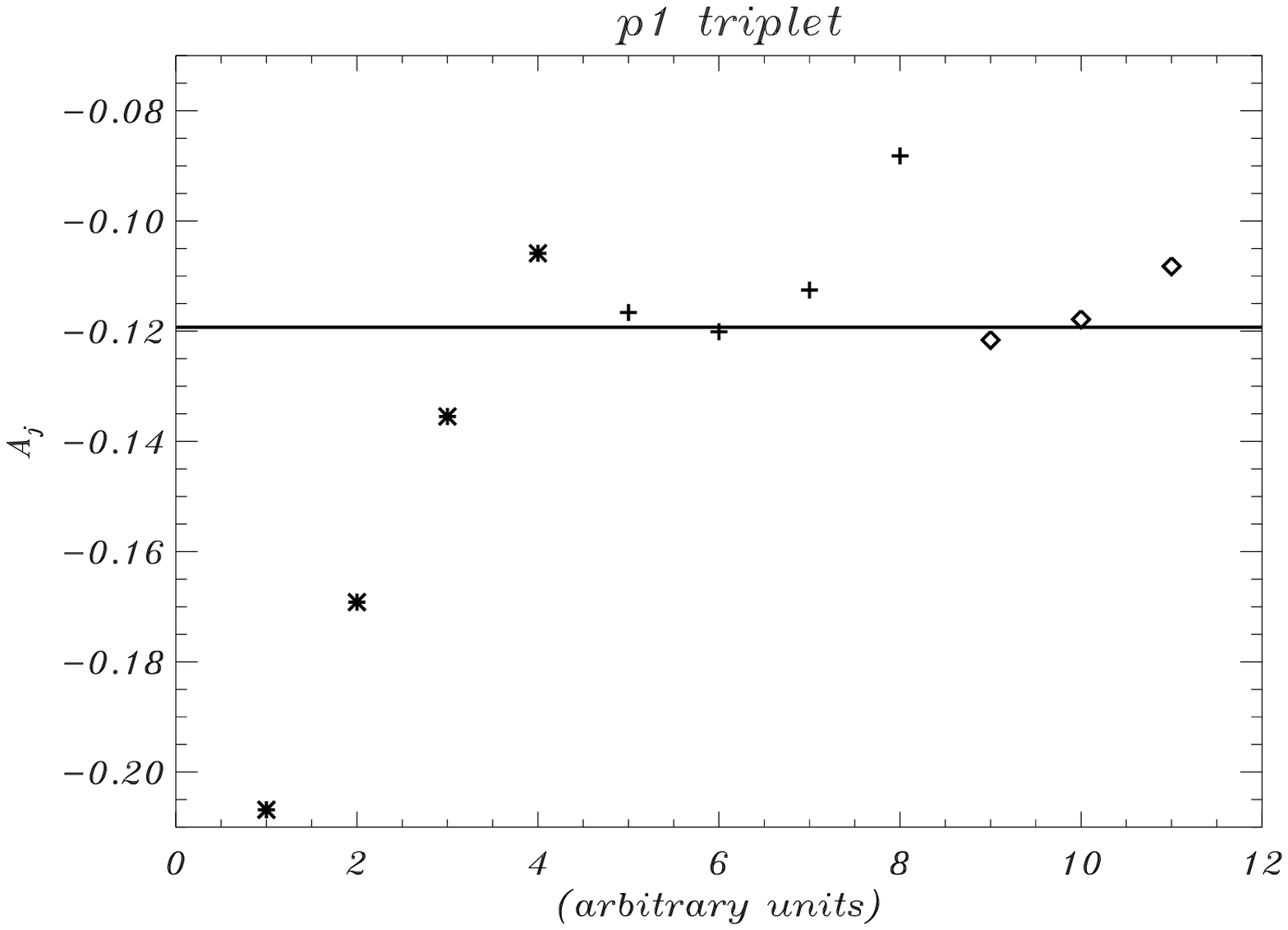}
    \caption{Absolute differences $\Delta K$ (top panels) between the kernels calculated for linear
             rotation profiles and the reference one (shellular profile), for the modes
             g1 (left), p1 (right). The $K$ subindices correspond the percentage of the core
             rotation rate of the reference model (see Fig.~\ref{fig:rotprof1})             
             Bottom panels show the theoretical asymmetries found
             for the different modes and rotation profiles. The horizontal line represents the
             splitting asymmetry of the g1 and p1 modes (left and right panels, respectively)
             found for the reference model. In order to identify the models we consider the 
             following representation: asterisks for linear profiles, crosses for smoothed 
             shellular rotation profiles, and rhombus for shellular rotation profiles.}
\label{fig:kerlinear_errors}
\end{figure*}

The study of the splitting asymmetry kernel is not an easy task. Structure and pulsation
variables, together with their derivatives, make the $K_2$ mathematical expression complex,
and hamper the interpretation of its variations in terms of physics. It is thus rather difficult to
invert the rotation profile by using the complete expression of $K_2$ (see appendix of Su\'arez et
al.~2006a). On the other hand, one expects that the \emph{pulsation terms} in the asymmetry kernel
are different for different mode types (pure p and g modes, and mixed modes). It is thus worth
studying the impact of variations in the rotation profile on the predicted asymmetries for the
different mode types. For the sake of simplicity, we will consider models typically representative
of \bceph\ stars.

To do so, we constructed $8\,\msun$ equilibrium models with the evolutionary code {\sc cesam}
(Morel 1997) for two different rotation profiles: 1) shellular rotation (we used this model as
the reference), obtained assuming local conservation of the angular momentum during the stellar
evolution, and 2) linear profiles, which drastically varies (with respect to reference model) the
rotational velocity in all the stellar interior. We forced all the models to have the same
rotational frequency at the surface (see Fig.~\ref{fig:rotprof1}). In this latter case, we computed
models with rotational frequency of the core $\nu=\Omega_c/\Omega_s= \nur, 0.75\,\nur, 0.50\,\nur$,
and $0.25\,\nur$,
respectively. 

We first analyzed the impact of the variations in the rotation profile on the splitting. We used
the oscillation code {\sc filou} (Su\'arez 2002, Su\'arez et al. 2008) to compute the oscillations.
Figure~\ref{fig:kerlinear_errors} (top panels) shows the kernel differences between the linear
rotation profiles models with the reference one $\Delta K_{j0}=K_j(r)-K_{\rm ref}(r)$ predicted for
g1 and p1 split modes. As expected, for both type of modes, the largest differences are 
located around the main peaks of the corresponding kernels (near the stellar core for g1, and
towards the stellar surface for p1). Even if p1 is a mixed mode (with energy in both the
surface and the core), $\Delta K_{j0}$ are larger for g1. This would help to disentangle (and
thereby identify) low frequency g and p modes, which are typically observed in \bceph\ stars.

We then calculated, for each triplet, the corresponding splitting asymmetry according to
Eq.~\ref{eq:defasym1} and compared them with the reference one (Fig.~\ref{fig:kerlinear_errors},
bottom panels). For illustration, we included in the figure the results for models
computed with \emph{modified} shellular-like rotation profiles (similar to the reference model but
modifying the rotational frequency of the core), and smoothed shellular-like rotation profiles, used
to discard numerical problems with derivatives near the convective core edge. 

A first glance indicates that the largest relative variations from the reference
asymmetry corresponds to the g1 triplet (up to 60\% ), and remains around
(up to 40\%) for the p1 triplet. As expected, the largest departs from the reference
asymmetry, $A_0$ corresponds to the linear rotation profile models (asterisks) which overestimate
it. In the case of smoothed shellular-like models, the results are very close to the
reference model for both modes with some marginal exception. Finally, the asymmetry found for 
the non-smoothed profiles (rhombus) varies significantly for g1, but remain very close to the
reference asymmetry value (p1 triplet).

Some preliminary conclusions can be extracted: rotational splitting asymmetries are
strongly sen\-si\-ti\-ve to chan\-ges of the rotation profile, and such a sensitivity is,
somehow,
differential, depending on the type of mode (g, p or mixed mode). Physically, the asymmetries
are dependent of second-order terms in the oscillation frequency that mainly account for the
distortion caused by the centrifugal force. However, we still need to understand how the structure
and pulsation variables are related to the asymmetry and its variations that
allow us to construct a simplified kernel for the splitting asymmetry (work in progress).
Moreover, thanks to the very precise data from space missions like \emph{CoRoT}
(Baglin et al. 2003), \emph{Kepler} (Gilliland et al. 2010), it is possible to measure with
unprecedent precision the rotational splittings and their asymmetries. This, together with
the increase of the number of detected modes (see e.g. Garc\'{\i}a Hern\'andez et al. 2009,
Poretti et al. 2009), will definitely help for the present study.

\acknowledgements
   JCS acknowledges support from the "Instituto de Astrof\'{\i}sica de Andaluc\'{\i}a (CSIC)" 
   by an "Excellence Project" post-doctoral fellowship, financed by the Spanish "Conjerer\'{\i}a de
   Innovaci\'on, Ciencia y Empresa de la Junta de Andaluc\'{\i}a" under proyect "FQM4156-2008".
   JCS also acknowledges support by the Spanish "Plan Nacional del Espacio" under project
   ESP2007-65480-C02-01.



\begin{thebibliography}{}

\bibitem{} Baglin, A.: 2003, Advances in Space Research, 31, 345

\bibitem{} {Bruntt}, H., {Stello}, D., {Su{\'a}rez}, J.~C., et al.: 2007a, MNRAS, 378, 1371

\bibitem{} {Bruntt}, H., {Su{\'a}rez}, J.~C., T.~R. {Bedding}, et al.: 2007b, A\&A, 461, 619

\bibitem{} {Casas}, R., {Su{\'a}rez}, J.~C., {Moya}, A., and {Garrido}, R.: 2006, A\&A, 455, 1019

\bibitem{} {Casas}, R., {Moya}, A., {Su{\'a}rez}, J.~C., et al.: 2009, ApJ, 697, 522

\bibitem{} Dziembowski, W.~A., \& Goode, P.~R.: 1992, ApJ, 394, 670 

\bibitem{} {Fox Machado}, L., {P{\'e}rez Hern{\'a}ndez}, F., {Su{\'a}rez}, J.~C., et al.: 2006
  A\&A, 446, 611

\bibitem{} Garc{\'{\i}}a Hern{\'a}ndez, A., {Moya}, A., {Michel}, E., {et~al.}: 2009,
  A\&A, 506, 79

\bibitem{} {Gilliland}, R.~L., {Brown}, T.~M., {Christensen-Dalsgaard}, J., {et~al.}: 2010,
  PASP, 122, 131

\bibitem{} Goupil, M.-J., Dupret, M.~A., Samadi, et al.: 2005, JApA, 26, 249 

\bibitem{} {Goupil}, M.~J. \& {Talon}, S.: 2009, CoAst 158, 220

\bibitem{} Goupil, M.~J.: 2009, LNP, Berlin Springer Verlag, 765, 45 

\bibitem{} Ligni{\`e}res, F., {Rieutord}, M., \& {Reese}, D.: 2006, A\&A, 455, 607

\bibitem{} {Morel}, P.: 1997, A\&AS, 124, 597

\bibitem{} {Moya}, A., J.~C. {Su{\' a}rez}, {Amado}, P.~J., et al.: 2005, A\&A, 432, 189

\bibitem{} Poretti, E., Michel, E., Garrido, R., et al.\ 2009, A\&A, 506, 85 

\bibitem{} {Reese}, D., {Ligni{\`e}res}, F., \& {Rieutord}, M. 2006, A\&A, 455, 621

\bibitem{} {Rodr{\'{\i}}guez}, E., {Costa}, V., {Zhou}, A.~Y., et al.: 2006a, A\&A, 456, 261

\bibitem{} {Rodr{\'{\i}}guez}, E., {Amado}, P.~J., {Su{\'a}rez}, J.~C., et al.: 2006b, A\&A 450, 715

\bibitem{} Soufi, F., Goupil, M.~J., \& Dziembowski, W.~A.: 1998, A\&A, 334, 911 
 
\bibitem{} {Su{\' a}rez}, J.~C.: 2002, Ph.D.~Thesis, ISBN 84-689-3851-3, ID 02/PA07/7178

\bibitem{} {Su{\' a}rez}, J.~C., {Bruntt}, H., \& {Buzasi}, D.: 2005a, A\&A, 438, 633

\bibitem{} {Su{\'a}rez}, J.~C., {Moya}, A., {Mart{\'{\i}}n-Ruiz}, S., et al.: 2005b, A\&A, 443, 271

\bibitem{} {Su{\'a}rez}, J.~C., {Goupil}, M.~J., and {Morel}, P.: 2006a, A\&A, 449, 673.

\bibitem{} {Su{\'a}rez}, J.~C., {Garrido}, R., \& {Goupil}, M.~J.: 2006b, A\&A, 447, 649

\bibitem{} {Su{\'a}rez}, J.~C., {Michel}, E., {Houdek}, G., et al.: 2007a, MRNAS, 379, 201

\bibitem{} {Su{\'a}rez}, J.~C., {Garrido}, R., \& {Moya}, A.: 2007b, A\&A, 474, 961

\bibitem{} {Su{\'a}rez}, J.~C., \& {Goupil}, M.~J.: 2008, Ap\&SS, 316, 155

\bibitem{} {Su{\'a}rez}, J.~C., {Moya}, A., {Amado}, P.~J., et al.: 2009, ApJ, 690, 1401

\bibitem{} K.~{Uytterhoeven}, P.~{Mathias}, {Poretti}, E., et al.: 2008, A\&A, 489, 1213

\bibitem{} Thompson, M.~J., {Toomre}, J., {Anderson}, E.~R.et al.: 1996, Science, 272, 1300 


\end{thebibliography}
\end{document}